\documentstyle[aps,preprint,epsfig]{revtex}

\title{\bf
The double parton distributions in the hard  pomeron model}
\author{Mikhail\, Braun}
\address{ Department of High-Energy Physics, University of St. Petersburg,
198904 St.Petersburg, Russia}
\author{Daniele\, Treleani}

\address{ Department of Theoretical Physics, University of Trieste, Strada
Costiera 11, Istituto Nazionale di Fisica Nucleare, Sezione di Trieste,
and ICTP, I--34014, Trieste,Italy}

\def\beq{\begin{equation}}
\def\eeq{\end{equation}}
\def\bea{\begin{eqnarray}}
\def\eea{\end{eqnarray}}

\begin{document}
\maketitle
\medskip
\vspace{1 cm}
{\bf Abstract.}
\vskip.2in
We study the double parton interaction process in collisions 
between highly virtual $q{\bar q}$ pairs in the BFKL regime. Explicit
expressions for the double parton distributions are obtained both in the case 
of direct coupling of the BFKL
pomerons to the $q{\bar q}$ pair and in the case of triple pomeron
interaction. 
\vspace{3cm}

E-mail braun1@MB1693.spb.edu \\

E-mail daniel@trieste.infn.it \\

\newpage
\section{ Introduction.}
An important feature of 
hadronic collisions at high energies is the growth of the hard component
of the interaction, induced by the increasing flux of partons.
Apart from the single hard interaction,
multi-parton hard interactions begin to play an increasing role.
Hard events
with multi-parton
interactions have in fact been predicted long ago by several
authors\cite{double}. 
The simplest event of this kind, the double parton
scattering, has been an object of the
experimental search in all high energy hadron collider
experiments since several years\cite{exdouble} and while initially the results
have been sparse and 
not very consistent, recently CDF has reported the observation
of a large number of events with double parton collisions\cite{cdf}. 

Multiple
parton collisions represent  a new observable feature of 
hadronic interactions.
Non-perturbative inputs to the corresponding cross sections are the
multi-parton distributions. These are new
properties of the hadron structure which become
accessible through the observation of multiple parton collisions.
The multiple
parton distributions are 
related directly to the many-body parton correlations in the
hadron\cite{ctcorr}. While, as a general rule, the non-perturbative inputs
are quantities to be measured and which cannot be computed in
perturbation
theory, in the interaction of two very virtual $q{\bar q}$ pairs, the
whole process falls in
the domain of perturbative QCD and, in that case, multiple parton
distributions can be
obtained through a direct calculation, within an approximation scheme.
The aim of the present paper is to study precisely this example of high
energy
interaction and to work
out explicitly the simplest case of multi-parton distribution. 

Note that studying events with several hard interactions one can
introduce standard notions of inclusive and exclusive cross-sections.
The first refer to the observation of at least one (or two,..) hard events,
the second of exactly one (or two,...) hard events. The inclusive
hard cross-sections are most easily accessable in the experimental
study.  They are also most easily calculated theoretically. 
In fact the inclusive cross sections are basically average quantities, 
since they include the multiplicity factor, which is proportional to the number of
hard partonic interactions\cite{ametller}, so no wonder that the
inclusive cross-section may become larger than the total
cross-section \cite{pancheri}.
The AGK rules\cite{agk} tell that the inclusive cross-sections for single,
double, triple etc hard collisions are given by the simplest
relevant diagrams, which involve
one, two, three etc hard interaction blobs. All other contributions
from diagrams with more hard blobs cancel in the
inclusive cross-sections. 
In the following, applying the simple theoretical formulas for the
inclusive cross-sections, we shall use them as basic quantities,
from which we shall extract the multiparton distributions.

The paper is organized in two parts. For the sake of completeness the
kinematics
of the single and double parton collision and
the factorization of the non-perturbative parts of the processes are
re-derived
in the first part of the paper. In the second part,
using the perturbative QCD approach, we compute
the double parton distribution in two different
limiting cases. The last paragraph is devoted to the conclusions.

\section{Single and double parton interactions}
\subsection{Single hard scattering}
To fix our notations and to settle a common ground with the double parton
collision process, we discuss first the single hard scattering case
which is described by the diagram shown in Fig.1.
The discontinuity of the diagram through the hard blob 
represents the inclusive cross section and, because of the AGK
cancellation\cite{agk}, it includes 
the factor accounting for the multiplicity of the hard
partonic collisions in the hadronic interaction\cite{ametller}. 
In the following one considers only
one sort of partons which, for the hard pomeron model,
is represented by the gluons.
The contribution of the diagram to the inclusive cross-section is related to
the amplitude by
\beq
\sigma=(1/s){\rm Im}{\cal A}
\eeq
where $s=2p_{+}q_{-}$ is the c.m energy squared. We neglect the
masses of the projectile and target and take
$p_{-}=q_{+}=p_{\perp}=q_{\perp}=0$. For massive projectile and target
the standard substitution $p\rightarrow p-\alpha q$, $q\rightarrow q-\alpha p$
with $\alpha=m^2/s$ is assumed. The usual scaling variables of
the colliding partons
are introduces as
\beq
k_{+}=xp_{+},\ \ l_{-}=wq_-
\eeq
The partonic c.m. energy squared is therefore written as 
$s_p\equiv M^2=(k+l)^2\simeq 2k_+l_-=sxw$. 

The standard treatment is based on two assumptions
which are the basis for the factorization hypothesis

* Parton virtualities (essentially the transverse momenta squared) are much
smaller as compared with $M^2$.

** The dependence of the forward amplitude $a$ on the
virtualities of the external lines can be neglected.

\begin{center}
  \begin{figure}
    \includegraphics[width=0.6\textwidth,clip]{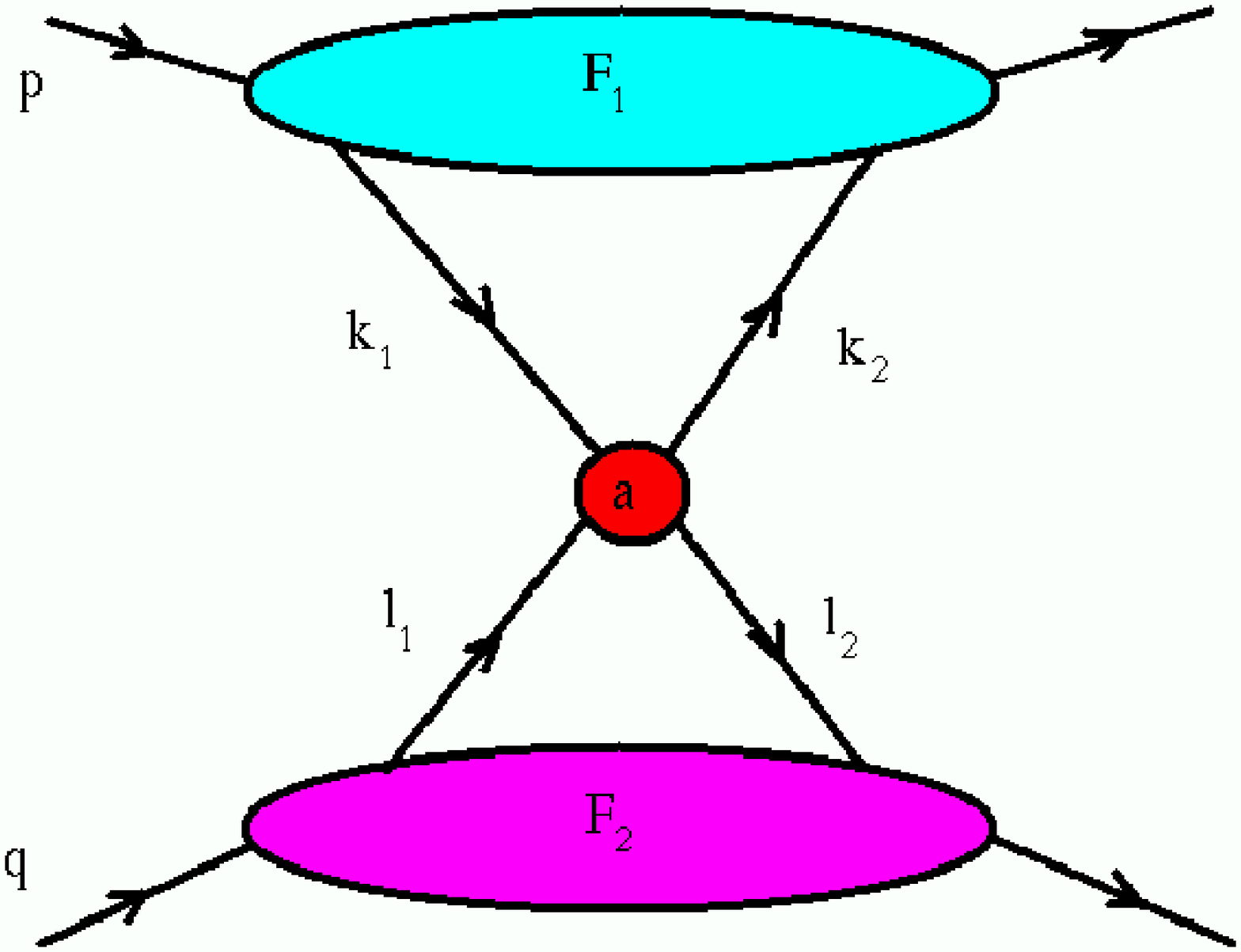}
    \caption{Single scattering term}
    \label
{br1}
  \end{figure}
\end{center}

The longitudinal integrations in the diagram
of Fig.\ref{br1} are easily done. Indeed according to our assumptions the amplitude $a$ does
not depend on $k_-$ nor on $l_+$. Only the upper part of the diagram depends
therefore on
$k_-$.
One denotes it (with parton legs) as $F_1(p,k)$. Its integration over
$k_-$ is standardly transformed into the integration over the
``missing mass'' $s_1=(p-k)^2$ of its right-hand discontinuity:
\beq
\int\frac{dk_-}{2\pi}F_1(p,k)=\frac{i}{2p_+}\int_{s_{10}}^{\infty}
\frac{ds_1}{\pi}{\rm Im}\,F_1(p,k)\equiv
\frac{2\pi i}{xp_+}F_1(x,k_{\perp})
\eeq
Similarly for the target
\beq
\int\frac{dl_+}{2\pi}F_2(q,l)=\frac{i}{2q_-}\int_{s_{20}}^{\infty}
\frac{ds_2}{\pi}{\rm Im}\,F_2(q,l)\equiv
\frac{2\pi i}{wq_-}F_2(w,l_{\perp})
\eeq
 
\begin{center}
  \begin{figure}
    \includegraphics[width=0.6\textwidth,clip]{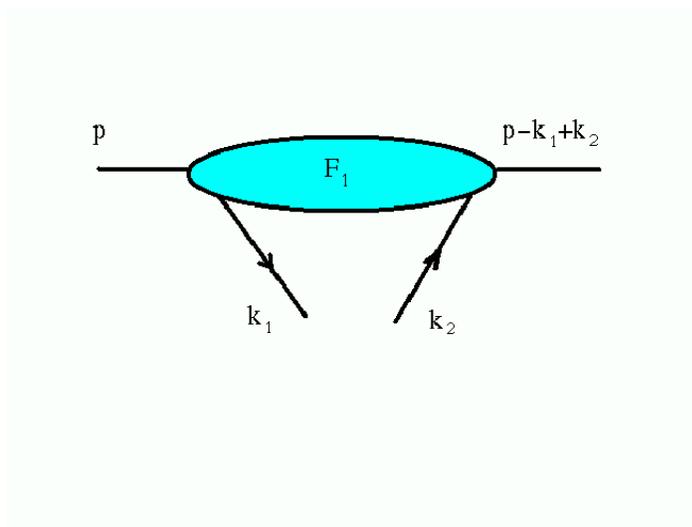}
    \caption{Soft blob in the single scattering term.}
    \label
{br2}
  \end{figure}
\end{center}

The standard parton distributions at the scale $M^2$ are obtained by integrating
$F_1$ and $F_2$ over the
transverse momenta $k_{\perp}$ and $l_{\perp}$:
\beq
\rho_1(x,M^2)=\int \frac{d^2k}{(2\pi)^2}\theta(M^2-k^2)F_1(x,k)
\eeq
The final integrations over $k_+$ and $l_-$ can be 
transformed into integrations over $x$ and $w$. The integration limits,
$[0,1]$, are a consequence
of the positivity constraints of
$s_1,s_2$ and $M^2$. The amplitude of Fig. 1 is therefore written as
\beq
{\cal A}_1=\int_0^1\frac{dxdw}{xw}\rho_1(x,M^2)\rho_2(w,M^2)a(M^2)
\eeq
while the single parton scattering inclusive cross-sections is expressed by the usual
factorized formula:
\beq
\sigma_1=\int_0^1 dxdw\rho_1(x,M^2)\rho_2(w,M^2)\sigma_p(xw)
\eeq
The cross section can be expressed by using the unintegrated parton
distributions in coordinate space. To this purpose one introduces
the (non-forward) target parton distribution  
$F_1(x,k_{1\perp},k_{2\perp})$, shown in Fig.\ref{br2}, and its Fourier transform
\beq
F_1(x,r_1,r_2)=\int\frac{d^2k_1d^2k_2}{(2\pi)^4}F_1(x,k_1,k_2)
e^{ik_1r_1-ik_2r_2}
\eeq
(In the following part of this subsection one will not find 4-vectors, so
we suppress the subindex $\perp$ for the $k$'s).
It is convenient to introduce the partonic c.m. and relative coordinates by
\beq
r_1+r_2=2R,\ \ r_1-r_2=r
\eeq
One has therefore
\beq
F_1(x,k_1,k_2)=\int d^2Rd^2rF_1(x,R,r)e^{-iR(k_1-k_2)-ir(k_1+k_2)/2}
\eeq
and when $k_1$ and $k_2$ are equal
\beq
F_1(x,k)=F_1(x,k,k)=\int d^2Rd^2rF_1(x,R,r)e^{-irk}
\eeq
which is precisely the unintegrated parton distribution which appeared in the
single hard scattering expression above. According to (5) one has to integrate
$F_1$ over all $k$ values, with the condition $k^2<<M^2$, in order to obtain the
final partonic distribution at the scale $M^2$. In the scaling
approximation the $\rho$'s do not depend on $M^2$. One may therefore take the limit
$M^2\rightarrow\infty$ and thus integrate over all $k$-values.
In the scaling approximation one therefore finds
\beq
\rho_1(x,M^2\rightarrow\infty)=\int d^2RF_1(x,R,0)
\eeq
When keeping into account the limits for
$k^2$ the value 
$r=0$ is changed into $r^2\simeq 1/M^2$.
The standard parton distributions $\rho(x,M^2)$, entering in the factorization
formula, can be therefore expressed via the density in
coordinate space $F$ as
\beq
\rho(x,M^2)=\int d^2RF(x,R,r), \ \ r^2=1/M^2
\eeq

\subsection{Double hard scattering}
\begin{center}
  \begin{figure}
    \includegraphics[width=0.6\textwidth,clip]{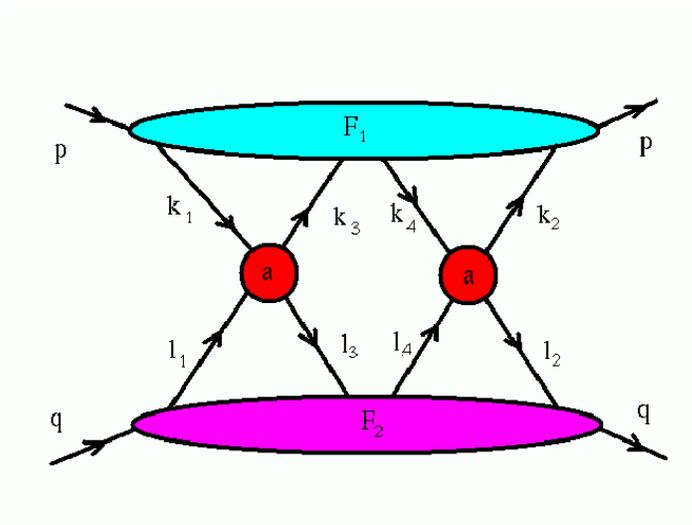}
    \caption{Double scattering term.}
    \label
{br3}
  \end{figure}
\end{center}

The
double hard scattering diagram is shown in Fig.\ref{br3}. The corresponding inclusive
cross section (including the multiplicity factor) is obtained, by means of the
AGK rules\cite{agk}, by taking twice the imaginary part and dividing it by
$-s$. 

The previous analysis can be repeated also in this case. There are
five double longitudinal integrations. As 
independent variables one chooses $k_1,k_2,l_1,l_2$ and $q=k_1-k_3$. The two hard scattering amplitudes are assumed to depend 
only on their total c.m. energies squared $M_1^2=(k_1+l_1)^2=2k_{1+}l_{1-}$ 
and $M_2^2=(k_2+l_2)^2=2k_{2+}l_{2-}$, not on
$k_{1(2)-}$, $l_{1(2)+}$ nor on $q_{\pm}$. The integrations on $k_{1(2)-}$ are
made as in the single scattering case.
One calls the projectile blob (with all four partonic legs)
$F_1(p,k_1,k_2,k_3,k_4)$ ($\sum k_i=0$). Then the integrations over $k_{1(2)-}$
can be transformed into integrations over the missing masses of the two
hard scatterings $s_1=(p-k_1)^2$ and $s_2=(p-k_2)^2$, the integrand being the
discontinuity of $F$ on the corresponding right-hand cuts.

Since neither the hard scattering amplitudes nor 
$F_2$ depend on $q_-$ one may integrate $F_1$ also over $q_-$. This 
last integration  can be transformed into an integration over the missing mass
in between the scatterings $s_3=(p-q)^2$ along the right-hand cut, the
integrand being the corresponding discontinuity. As a result one obtains
\[
\left(\frac{i}{2p_+}\right)^3\int_{s_{10}}^{\infty}\frac{ds_1}{2\pi i}
\int_{s_{20}}^{\infty}\frac{ds_2}{2\pi i}
\int_{s_{30}}^{\infty}\frac{ds_3}{2\pi i}{\rm Disc}_{s_1}
{\rm Disc}_{s_2}{\rm Disc}_{s_3}F_1(p,k_1,k_2,k_3,k_4)\]\beq
\equiv
-2\pi^2 i\left(\frac{1}{p_+}\right)^3
F_1(x_1,k_{1\perp},k_{3\perp}|x_2,k_{4,\perp},k_{2\perp})
\eeq
The density  defined on the right-hand side 
is the generalization of the single
to the double unintegrated partonic density.
The double parton distribution of the target is introduced in an analogous way.
The integrations over $k_{1(2)+}$, $l_{1(2)-}$ are transformed into
integrations on the scaling variables $x_{1,2}$ and $w_{1,2}$ and the
double hard scattering amplitude of Fig.\ref{br3} is expressed as
\[
{\cal A}_2=\frac{i}{2s}\int_0^1\frac{dx_1}{ x_1}\frac{dx_2}
{x_2}\frac{dw_1}{w_1}
\frac{dw_2}{ w_2}\int\prod_{j=1}^4\frac{d^2k_j}{(2\pi)^2}
\frac{d^2l_j}{(2\pi)^2}(2\pi)^2\]\[
\delta^2(k_1-k_2-k_3+k_4)
(2\pi)^2\delta^2(k_1+l_1-k_3-l_3)(2\pi)^2\delta^2(k_2+l_2-k_4-l_4)
\]\beq
F_1(x_1,k_{1},k_{3}|x_2,k_{4},k_{2})
F_2(x_2,l_{1},l_{3}|x_2,l_{4},l_{2})a(M_1^2)a(M_2^2)
\eeq
where $M_1^2=sx_1w_1$ and $M_2^2=sx_2w_2$. The double hard scattering cross
section is therefore written as
\[
\sigma_2=\int_0^1dx_1dx_2dw_1dw_2
\int\prod_{j=1}^4\frac{d^2k_j}{(2\pi)^2}\frac{d^2l_j}
 {(2\pi)^2}\]\[(2\pi)^2\delta^2(k_1-k_2-k_3+k_4)
(2\pi)^2\delta^2(k_1+l_1-k_3-l_3)(2\pi)^2\delta^2(k_2+l_2-k_4-l_4)
\]\beq
F_1(x_1,k_{1},k_{3}|x_2,k_{4},k_{2})
F_2(w_1,l_{1},l_{3}|w_2,l_{4},l_{2})
\sigma_p(x_1w_1)\sigma_p(x_2w_2)
\eeq
and both in Eq.(15) and in Eq.(16) the integrations are done with the limits
$k_1^2,k_3^2<M_1^2$ and
$k_2^2,k_4^2<M_2^2$.

The integrations over the transverse components of the transferred momenta are
more conveniently performed by going to coordinates space. One introduces therefore
\beq
F_1(x_1,r_1,r_3|x_2,r_4,r_2)=
\int\prod_{j=1}^4\frac{dk_j}{(2\pi)^2}
F_1(x_1,k_{1},k_{3}|x_2,k_{4},k_{2})
e^{(ik_1r_1-ik_2r_2-ik_3r_3+ik_4r_4)}
\eeq
where in the integrand all four momenta are independent, in such a way that
one is integrating also over the 
total  momentum transferred to the projectile. One introduces the c.m. and
relative coordinates defined as:
\beq
r_1+r_3=2R_1,\ \ r_1-r_3=r_{13},\ \ r_4+r_2=2R_2,\ \ r_2-r_4=r_{24}
\eeq
so that
\[ r_1=R_1+r_{13}/2,\ \  r_3=R_1-r_{13}/2,\ \  r_2=R_2+r_{24}/2, \ \ 
r_4=R_2-r_{24}/2\]
Similar coordinates, with primes, are introduced for the target:
\[ l_1+l_3=2R'_1\ \ {\rm etc.}\]
Going to the coordinates space one finds, in Eq.(16), the following exponential
\[
\exp\Big\{-iR_1(k_1-k_3)-ir_{13}(k_1+k_3)/2+iR_2(k_2-k_4)+ir_{24}(k_2+k_4)/2\]\[
-iR'_1(l_1-l_3)-ir'_{13}(l_1+l_3)/2+iR'_2(l_2-l_4)+ir'_{24}(l_2+l_4)/2\Big\}\]
Also the three $\delta$ functions in Eq.(16) can be represented as integrals
over three additional coordinates in transverse space:
\[
(2\pi)^2\delta^2(k_1-k_2-k_3+k_4)=\int d^2Re^{iR(k_1-k_2-k_3+k_4)}
\]
\[
(2\pi)^2\delta^2(k_1+l_1-k_3-l_3)=\int d^2b_1e^{ib_1(k_1+l_1-k_3-l_3)}
\]
\[
(2\pi)^2\delta^2(k_4+l_4-k_2-l_2)=\int d^2b_2e^{ib_2(k_4+l_4-k_2-l_2)}
\]
In the scaling limit the integrations over the eight
momenta $k_j,l_j$, $j=1,...4$ produce eight $\delta$
functions:
\[\delta^2(R+b_1-R_1-r_{13}),\ \
\delta^2(-R-b_2+R_2+r_{24}),\ \ 
\delta^2(-R-b_1+R_1-r_{13}),\]\[
\delta^2(R+b_2-R_2+r_{24}),\ \ 
\delta^2(b_1-R'_1-r'_{13}),\ \ 
\delta^2(-b_2+R'_2+r'_{24}),\]\[
\delta^2(-b_1+R'_{1}-r'_{13}),\ \ 
\delta^2(b_2-R'_{2}+r'_{24})\]
The integrations over the transverse coordinates give then $r_{13}=r'_{13}=r_{24}=r'_{24}=0$,
$R'_{1}=b_1$, $R'_2=b_2$, $R_1=R+b_1$, $R_2=R+b_2$. 
As a result the double hard cross-section is expressed as an
integral over the three impact parameters, $R, b_1$ and $b_2$
\[
\sigma_2=\int_0^1dx_1dx_2dw_1dw_2
\sigma_p(x_1w_1)\sigma_p(x_2w_2)\]\beq\int d^2Rd^2b_1d^2b_2
F_1(x_1,R+b_1,0|x_2,R+b_2,0)
F_2(w_1,b_1,0|w_2,b_2,0)
\eeq

As in the case of the single parton scattering
process, to keep into account that the integrations on the transverse momenta
cannot be extended to infinity, one takes the relative transverse distances
between the interacting partons to be of order of the dimensions of the two
hard scattering blobs, namely is $1/M_1$ and $1/M_2$. The resulting 
expression of the double parton scattering cross section is therefore
written as
\[
\sigma_2=\int_0^1dx_1dx_2dw_1dw_2
\sigma_p(x_1w_1)\sigma_p(x_2w_2)\int d^2Rd^2b_1d^2b_2\]\beq
F_1(x_1,R+b_1,r_{13}|x_2,R+b_2,r_{24})
F_2(w_1,b_1,r_{13}|w_2,b_2,r_{24})
\eeq
with $r^2_{13}=1/M_1^2$ and $r^2_{24}=1/M_2^2$.

\begin{center}
  \begin{figure}
    \includegraphics[width=0.6\textwidth,clip]{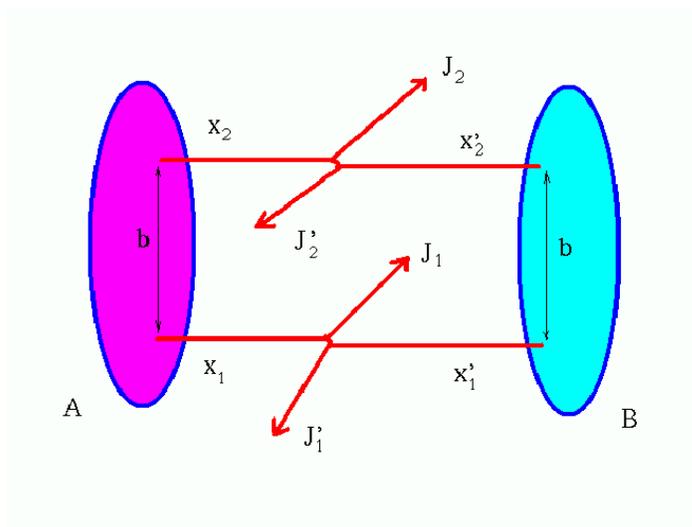}
    \caption{Graphical representation of Eq.\ref{sigmad}.}
    \label
{doubleb}
  \end{figure}
\end{center}

The expression is further simplified by introducing the c.m. and relative
parton coordinates in transverse space

\[
B_1=(b_1+b_2)/2+R,\  \ B_2=(b_1+b_2)/2,\  \ b=b_1-b_2
\]
The double parton distributions are therefore defined as
\[
\Gamma_1(x_1,x_2,b;M_1^2,M_2^2)=\int
d^2B_1F_1(x_1,R+b_1,r_{13}|x_2,R+b_2,r_{24})\]\beq
\Gamma_2(w_1,w_2,b;M_1^2,M_2^2)=\int d^2B_2F_2(w_1,b_1,r_{13}|w_2,b_2,r_{24})
\eeq
and the double parton scattering cross section assumes a simpler form
\beq
\sigma_2=\int\Gamma_1(x_1,x_2,b;M_1^2,M_2^2)\sigma_p(x_1w_1)
\sigma_p(x_2w_2)\Gamma_2(w_1,w_2,b;M_1^2,M_2^2)dx_1dw_1dx_2dw_2d^2b
\label{sigmad}
\eeq
with the geometrical interpretation shown
in fig.\ref{doubleb}.
Notice that the normalization contains also the multiplicity factor which
counts the number of parton interactions in the hadronic collision (two in
this case).  

\section{Double parton distributions in the BFKL regime}
\subsection{Emission of a jet from a single BFKL chain}
In the framework of the hard pomeron model the hard collision process is
associated to the  production of a gluon with sufficiently large 
transverse momentum $k>k_{min}$ and at a given rapidity (In this section only 2-dimensional transverse
vectors will appear, so that we again omit the subindex $\perp$ and one take
the
metric Euclidean, $k^2\geq 0$). 

The emission of a gluon with momentum $k$ is
described by modifying as follows the BFKL Green function in coordinates space
$G(r',r'')$\cite{bt}:
\beq
G(r',r'')\rightarrow \int d^2rG(r'r)V(k)G(r,r'')
\eeq
where $r$'s are relative distances between the gluons and
\beq
V_k(r)=\frac{12\alpha_s}{k^2}\stackrel{\leftarrow}{\Delta} e^{ikr}
\stackrel{\rightarrow}{\Delta}
\eeq
is the vertex operator describing the emission of the gluon,
the arrows show the direction of the
action of the $\Delta$ operators. 
The two BFKL Green functions in (23) are to be taken at
appropriate energy (rapidity) ranges, corresponding to the 
rapidity distance of the emitted gluon from the projectile. This dependence 
is not written explicitly nor to overburden our formulas  with arguments and/or indices.

The first step is to generalize Eq.(23) for the non-forward BFKL chain. In momentum space
the emission of a gluon with momentum $k$ is described by the vertex
\begin{eqnarray}
V(l_1,l_2|l'_1,l'_2)&&\nonumber\\
=&&\frac{6\alpha_s}{k^2}
(2\pi)^2\delta^2
\left(\frac{1}{2}(l_1+l_2-l'_1-l'_2)-k\right)
\Big(l_1^2{l'_2}^2+l_2^2{l'_1}^2-{k^2}(l_1-l_2)^2\Big)
\end{eqnarray}
So the emission blob in momentum space $Z_k(q_1,q_2|q'_1,q'_2)$ is
given by
\begin{eqnarray}
Z_k(q_1,q_2|q'_1,q'_2)&=&\frac{6\alpha_s}{k^2}\int\prod_{i=1}^2
\frac{d^2l_i}{(2\pi)^2}\frac{d^2l'_i}{(2\pi)^2}
(2\pi)^2\delta^2
\left(\frac{1}{2}(l_1+l_2-l'_1-l'_2)-k\right)
\nonumber\\
&\times&
\Big(l_1^2{l'_2}^2+l_2^2{l'_1}^2-{k^2}(l_1-l_2)^2\Big)
G(q_1,q_2|l_1,l_2)G(l'_1,l'_2|q'_1q'_2)\nonumber\\
&\times&(2\pi)^2\delta^2(q_1-q_2-l_1+l_2)
(2\pi)^2\delta^2(l'_1-l'_2-q'_1+q'_2)
\end{eqnarray}
In coordinates space one writes:
\begin{eqnarray}
(2\pi)^2\delta^2(q_1-q_2-l_1&+&l_2)G(q_1,q_2|l_1,l_2)\nonumber\\
&=&\int \prod_{i=1}^2d^2r_id^2z_iG(r_1,r_2|z_1,z_2)
e^{(-iq_1r_1+iq_2r_2+il_1z_1-il_2z_2)}
\end{eqnarray}
and
\begin{eqnarray}
(2\pi)^2\delta^2(l'_1-l'_2-q'_1&+&q'_2)G(l'_1,l'_2|q'_1,q'_2)\nonumber\\
&=&
\int \prod_{i=1}^2d^2r'_id^2z'_iG(z'_1,z'_2|r'_1,r'_2)
e^{(-il'_1z'_1+il'_2z'_2+iq'_1r'_1-iq'_2r'_2)}
\end{eqnarray}
Note that due to the inclusion of the $\delta$ function in the Fourier
transforms (25) and (26), the Green functions in coordinate space
are translationally invariant:
\beq
G(r_1+a,r_2+a|z_1+a,z_2+a)=G(r_1,r_2|z_1,z_2)
\eeq 
The momenta squared in the emission vertex are substituted by
differential
operators applied to the Green function. The bracket in (24) 
is written as
\[\Delta_1\Delta'_2+\Delta_2\Delta'_1+
k^2(\nabla_1+\nabla_2)(\nabla'_1+\nabla'_2)\]
where one uses the notation $\Delta_1=\Delta(z_1)$, $\Delta'_1=\Delta(z'_1)$
etc.
The remaining $\delta$ function in Eq.(24) is represented as an integral
\[
(2\pi)^2\delta^2((l_1+l_2-l'_1-l'_2)/2-k)=\int d^2z
e^{-iz((l_1+l_2-l'_1-l'_2)/2-k)}
\]
in such a way that the integrations over the four transverse momenta in Eq.(24) produce four
$\delta$ functions in coordinate space:
\[\delta^2(z_1-z/2),\ \ \delta^2(z_2+z/2),\ \
\delta^2(z'_1-z/2),\ \ \delta^2(z'_2+z/2)\]
One obtains therefore
\[
Z_k(q_1,q_2|q'_1,q'_2)=\int\prod_{i=1}^2
d^2r_id^2r'_ie^{(-iq_1r_1+iq_2r_2+iq'_1r'_1-iq'_2r'_2)}
d^2z\Big[G(r_1,r_2|z_1,z_2)\frac{6\alpha_s}{k^2}e^{ikz}\]\beq
\times(\Delta_1\Delta'_2+\Delta_2\Delta'_1+
k^2(\nabla_1+\nabla_2)(\nabla'_1+\nabla'_2))
G(z'_1,z'_2|r'_1,r'_2)\Big]_{z_1=z'_1=\frac{z}{2},z_2=z'_2=-\frac{z}{2}}
\eeq 
and the differential operator which stands between the two Green functions
is precisely the generalization of the emission operator $V_k$ to the
non-forward case. By using the notation:
\[
G\Big(r_1,r_2\Big|\frac{z}{2},-\frac{z}{2}\Big)V_k(z)G
\Big(\frac{z}{2},-\frac{z}{2}\Big|r'_1,r'_2\Big)\equiv
\Big[G(r_1,r_2|z_1,z_2)\frac{6\alpha_s}{k^2}e^{ikz}\]\beq
\times(\Delta_1\Delta'_2+\Delta_2\Delta'_1+
k^2(\nabla_1+\nabla_2)(\nabla'_1+\nabla'_2))
G(z'_1,z'_2|r'_1,r'_2)\Big]_{z_1=z'_1=\frac{z}{2},z_2=z'_2=-\frac{z}{2}}
\eeq
one may rewrite Eq.(28) in a more compact form
\begin{eqnarray}
Z_k(q_1,q_2|q'_1,q'_2)=\int\prod_{i=1}^2&&
d^2r_id^2r'_ie^{(-iq_1r_1+iq_2r_2+iq'_1r'_1-iq'_2r'_2)}
d^2z\nonumber\\
&&\times G\Big(r_1,r_2\Big|\frac{z}{2},-\frac{z}{2}\Big)V_k(z)
G\Big(\frac{z}{2},-\frac{z}{2}\Big|r'_1,r'_2\Big)
\end{eqnarray}
One defines $Z$ in coordinates space in the following way:
\begin{eqnarray}
Z_k(r_1,r_2|r'_1,r'_2)&=&\int\prod_{j=1}^2\frac{d^2q_i}{(2\pi)^2}
\frac{d^2q'_j}{(2\pi)^2}\nonumber\\
&\times&(2\pi)^2\delta(q_1-q_2-q'_1+q'_2)
Z_k(q_1,q_2|q'_1,q'_2)
e^{(iq_1r_1-iq_2r_2-iq'_1r'_1+iq'_2r'_2)}
\end{eqnarray}
By expressing the $\delta$ function as an integral over the
impact parameters $B$
\beq
(2\pi)^2\delta(q_1-q_2-q'_1+q'_2)=
\int d^2Be^{-iB(q_1-q_2-q'_1+q'_2)}
\eeq
and by using  (32) and (34) in (33) one can do all integrations with the
exception of those on
$B$ and $z$. The result is
\begin{eqnarray}
Z_k(r_1,r_2|r'_1,r'_2)&&\nonumber\\
=\int d^2Bd^2&z&
G\Bigl(r_1+B,r_2+B\Big|\frac{z}{2},-\frac{z}{2}\Bigr)V_k(z)G\Bigl(\frac{z}{2},-\frac{z}{2}\Big|r'_1+B,r'_2+B\Bigr)
\end{eqnarray}
and, as a function of the c.m and relative coordinates,
\beq
Z_k(R,r|R',r')=\int d^2Bd^2z
G(R+B,r|0,z)V_k(z)G(0,z|R'+B,r')
\eeq
showing that $Z_k$ is translationally invariant:
\beq
Z_k(R,r|R',r')=Z_k(R+B,r|R'+B,r')=Z_k(R-R',r,r')
\eeq
Eq. (37) is our final expression. It shows how one should change a BFKL
Green function $G(R,r|R',r')=G(R-R',r,r')$ to describe the emission of a gluon
from the BFKL chain.

\subsection{Single parton distributions}
The simplest case to consider in the BFKL formalism is the inclusive emission 
of a gluon.
The problem has been extensively studied in the
literature\cite{pom}. Here we approach the problem from a 
somewhat novel point of view, introducing the single parton densities
in coordinates space in the BFKL framework. The quantities which are factorized
in the BFKL formalism are the `unintegrated gluon densities'\cite{lowx},
depending explicitly on the transverse momentum of the interacting gluon. The
distributions that will be discussed hereafter are therefore different as 
compared to the 
distributions discussed in the first part of the paper, since they depend
explicitly on the distance in transverse space between the initial and the final
interacting partons. The latter distributions are therefore obtained only after
integration on this transverse distance, with $1/k$ as a lower limit. 

Rather than using the known expression for
the inclusive cross-section in terms of forward BFKL Green functions,
one starts form the 
amplitude, corresponding to the exchange
of any number of pomerons between the projectile and target,
obtained in the approximation of a large number of colors $N$
and assuming the coupling of pomerons to the colliding particles
as purely perturbative\cite{b1}. The amplitude is written as
\beq
{\cal A}=2is 
\int\,d^{2}R \int d^{2}r \int d^{2}r'
\rho_p(r)\rho_q(r')
\Big(1-e^{-\frac{1}{2}g^4G_s(R,r,r')}\Big)
\eeq
Here $\rho_{p(q)}(r)$ are the color densities of the projectile and
of the target (with momenta $p$ and $q$ respectively). The function $G_s(R,r,r')=G(R,r|0,r')$
is the BFKL Green function, depending on the Mandelstam invariant $s$. When retaining the
$s$-wave only, because of the azimuthal symmetry, it takes the form\cite{lipatov}
\beq
G_s(R,r,r')=\frac{1}{(2\pi)^4}\int
d\nu\frac{\nu^{2}s^{-E(\nu)}}{(\nu^{2}+1/4)^{2}}
\int d^{2}r_{0}
\left(\frac{r}{r_{10}r_{20}}\right)^{1+2i\nu}
\left(\frac{r'}{r'_{10}r'_{20}}\right)^{1-2i\nu}
\eeq
where \[r_{10}=R+r/2-r_{0},\ r_{20}=R-r/2-r_{0},\ r'_{10}=r'/2-r_{0},\
r'_{20}=-r'/2-r_{0}\]
and $E(\nu)$ is the BFKL pomeron energy
\beq
E(\nu)=(3g^{2}/2\pi^{2})({\rm Re}\,\psi(1/2+i\nu)-\psi(1))
\eeq

Eq.(38) allows us to write down easily the amplitudes for single, double
etc. pomeron exchange processes. The single BFKL pomeron exchange amplitude is
written as
\beq
{\cal A}_1=isg^4\alpha_s^2 
\int\,d^{2}R \int d^{2}r \int d^{2}r'
\rho_p(r)\rho_q(r')
G_s(R,r,r')
\eeq
where the factor $g^4$ corresponds to the couplings of the pomeron to the
color densities $\rho$.
The inclusive cross-section for
emission of a single gluon has therefore the form
\[
I(k)=\frac{(2\pi)^2d^3\sigma}{dyd^2k}=
g^4\int\,d^{2}R \int d^{2}r \int d^{2}r'
\rho_p(r)\rho_q(r')
Z_{k}(R,r,r')\]\[=
\int\,d^{2}R \int d^{2}r \int d^{2}r'
g^2\rho_p(r)g^2\rho_q(r')
d^2Bd^2z
G_{s_1}(R+B,r|0,z)V_k(z)G_{s_2}(0,z|B,r')\]\beq
=\int\,d^{2}Rd^2R'd^2z \int d^{2}r \int d^{2}r'
g^2\rho_p(r)g^2\rho_q(r')
G_{s_1}(R,r|0,z)V_k(z)G_{s_2}(0,z|R',r')
\eeq
The dependence on energy is written explicitly in the Green
functions. The emitted gluon rapidity (relative to the target) is 
$y=\log s_2$. Then $s_1=s/s_2$ (the scale is not defined in the theory, as
usual one takes it as 1 GeV).
The integrated single hard scattering inclusive cross section can
be written as
\[\sigma_1=\int dy d^2k I(k)=\int d^2k\int_0^1dx dw\delta\left(xw-
\frac{k^2}{s}\right)I(k)\]
where one has introduced the scaling variables $x=s_1/s$ and $w=s_2/s$.
By comparing the expression above with the factorization formula
one obtains the expression for the partonic density of the projectile
\beq
F_p(x,R,z)=g^2\int d^2r\rho_p(r)G_{s_1}(R,r,z)
\eeq
with $x=s/s_1$ and $G(R,r,z)=G(R,r|0,z)$.

The meaning of the variables is the following: $R$ is the distance from the
center of the projectile and $z$ the distance between the
initial and the final interacting partons. The size of $z$ is of order of
$1/k$, as a result of the factor $\exp ikz$ in the hard interaction
vertex, and the size of 
$r$ is of the order of the dimensions of the projectile.
The perturbative approach is justified when both $r$ and $z$ are small. 
The asymptotics of the Green function $G$ for  small
relative distances and large energy was obtained in\cite{b1}. The resulting
expression is a function of the
dimensionless variable $\xi=R^{2}/rr'(>>1$):
\beq
G_s(R,r,r')_{r,r'<<R}\simeq \frac{s^{\Delta}}
{\pi^{5/2}(a\ln s)^{3/2}}\xi^{-1}\ln \xi\exp \left(-\frac{\ln^{2} \xi}
{a\ln s}\right)
\eeq
where $\Delta$ is the BFKL intercept
\beq
\Delta =\frac{12\alpha_s}{\pi}\ln 2
\eeq 
and 
\beq
a=\frac{42\alpha_s}{\pi}\zeta (3)
\eeq
With $z$ small and fixed, the partonic density (43) behaves as
\beq
F_p(x,R,z)\sim\frac{(1/x)^{\Delta}}{\log^{3/2}(1/x)}\frac{1}{R^2}
\log R\exp\left(-\frac{\log^2 R}{a\log(1/x)}\right)
\eeq
and it falls rather slowly (essentially as $1/R^2$)
as a function of the distance from the center of
the projectile ($R$ represents the distance from the projectile,
whose average dimension $r_0=\langle r\rangle$ is much
smaller than
$R$). The effect of the exponential is felt 
only at relatively large $R\sim\exp\sqrt{1/x}$. A more quantitative 
estimate
of the density can be made by neglecting the weak dependence on $r$
in the logarithmic factors in (44) and by making the substitution $r\rightarrow r_0$:
\beq
F_p(x,R,z)\simeq
\frac{g^2}{\pi^{5/2}}
\frac{(1/x)^{\Delta}}
{(a\ln (1/x))^{3/2}}\frac{zr_0}{R^2}
\ln \frac{R^2}{z r_0}\exp 
\left(-\frac{\ln^{2}(R^2/z r_0)}{a\ln (1/x)}\right)
\eeq
An analogous expression can be written for the target.

\subsection{Two pomerons coupling to a $q\bar{q}$ pair and triple pomeron}
\begin{center}
  \begin{figure}
    \includegraphics[width=0.6\textwidth,clip]{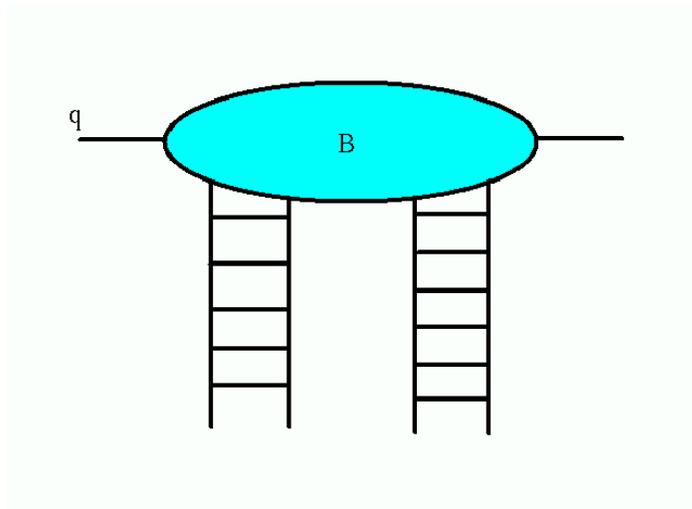}
    \caption{Double parton density in the case of direct coupling of the BFKL 
    pomerons to the
    constinuent quarks.}
    \label
{br4}
  \end{figure}
\end{center}

The general structure of
the amplitude describing the double partonic density in the BFKL regime is
shown in Fig.\ref{br4}, where one assigns the partons undergoing the two hard
collisions to two different BFKL pomeron ladders.
The coupling of
the two pomerons to the projectile is described by the upper blob $B$. 
One calls the energetic variable of the pomeron $s_2=s/s_1$, with $s$ the
overall c.m. energy squared, for simplicity 
one takes it to be the same for both legs. The upper blob $B$ is then integrated
on $s_1$.

The asymptotic behavior of $B(s_1)$ is described by a single pomeron
exchange, so that
\beq
B(s_1)\equiv B(y_1)=b(y_1)e^{\Delta y_1},\ \ y_1=\log s_1
\eeq
where $\Delta$ is the pomeron intercept and $b(y)$ is some smooth 
function tending to a constant at high $y_1$.
Each of the pomeron legs has a similar behavior as a function of its energy
variable $s_2$:
\beq
P(s_2)\equiv P(y_2)=p(y_2)e^{\Delta y_2},\ \ y_2=\log s_2
\eeq
with $p(y_2)$ a smooth function tending to a constant at high $y_2$.
The amplitude of Fig.\ref{br4} is given by a integral over $y_1$:
\beq
\int_0^Ydy_1B(y_1)P_1(Y-y_1)P_2(Y-y_1)=e^{2\Delta Y}
\int_0^Ydy_1e^{-\Delta y_1}b(y_1)p_1(Y-y_1)p_2(Y-y_1)
\eeq
where $Y=y_1+y_2=\log s$ is the overall rapidity interval. As one may see, 
at very large $Y$ values, the integration in (51) stays  limited to the
region
where $y_1$ is smaller or of the order $1/\Delta$ and thus it is independent on
$Y$. 
One may therefore take for $p_{1(2)}$ the asymptotic values and rewrite (51) as
\beq
e^{2\Delta Y}p_1^{as}p_2^{as}\int_0^\infty e^{-\Delta y_1}b(y_1)dy_1
\eeq
which corresponds precisely to the behavior of a two pomeron exchange 
process.

The rapidities that dominate the integral (52) are of order $1/\Delta$.
The actual value $\Delta$ is however not well defined, since
it is given by the coupling, whose value is a parameter in the BFKL
approach. Rather than trying to determine which are the rapidities relevant to
$B$, we proceed by
discussing two limiting cases.

If $Q^2$ is the virtuality of the interacting $q{\bar q}$ pairs (which has
to be large to justify the perturbative approach) a limiting
configuration is reached when $Q^2$ is of the same order (or larger) than $s_1$, namely,
\beq
\log Q^2\sim ({\rm or}\,>) 1/\Delta<<Y
\eeq
The blob $B$ enters then in the DGLAP regime (finite scaling variable $x_1$)
where only the large logarithms
$\log (Q^2/\Lambda)$ are to be considered. In the fixed coupling approach
of BFKL these logarithms are neglected altogether. Within the present
scale invariant model  one does not need therefore to sum any logarithms and
one may state within the pure perturbative approach. The upper blob is then
simply reduced
to the $q\bar{q}$ pair, to which the pomerons are coupled directly.
This is precisely the approximation which leads to the amplitude (36) obtained 
in\cite{b1}
and which was
used in the previous subsection.

On the other hand, at lower virtualities, such as
\beq
1<<\log Q^2<<1/\Delta
\eeq
the structure of the blob $B$ becomes important. Another limiting
configuration is reached when
$B$ enters in the BFKL regime, namely it is itself
described by a pomeron, which has to eventually split
into the two lower pomerons and the whole amplitude is
described by a triple pomeron interaction. One should keep in mind
that, in this last case, the three pomerons are not however in an equivalent
regime.
The two lower ones, as mentioned, may be taken at their asymptotical
regime, $\Delta y_2>>1$. The upper one, on the contrary, is characterized by
much smaller energies, such that $\Delta y_1\sim 1$, so that its exact form
needs to be used.

The two limiting possibilities which we consider are therefore the direct
coupling of the pomerons to the constituent quarks
and the triple pomeron interaction.
We first discuss the simpler direct coupling case.

\subsection{Two pomerons coupled directly to the projectile(target)}
From the general expression of
the amplitude (36) the contribution with two pomerons
directly coupled both to the projectile and target is written as:
\beq
{\cal A}_2=-is\frac{g^8}{2} 
\int\,d^{2}R \int d^{2}r \int d^{2}r'
\rho_p(r)\rho_q(r')
G_s^2(R,r,r')
\eeq
The integrated inclusive cross section is obtained 
by dividing the amplitude by $s$ and by taking 
the imaginary part with a minus sign:
\beq
\sigma_2=g^8 
\int\,d^{2}R \int d^{2}r \int d^{2}r'
\rho_p(r)\rho_q(r')
G_s^2(R,r,r')
\eeq
To obtain the double differential inclusive cross-section, corresponding to the emission
of two hard gluons with momenta $k_1$ and $k_2$ from the two pomerons, we
have to substitute the pomeronic Green function with the emission functions
$Z_{k_1}$ and $Z_{k_2}$. One obtains therefore
\[
I(k_1,k_2)=\frac{(2\pi)^2d^3}{dy_1d^2k_1}\frac{(2\pi)^2d^3}{dy_2d^2k_2}\sigma
\]\beq
g^8\int d^{2}R \int d^{2}r \int d^{2}r'
\rho_p(r)\rho_q(r')
Z_{k_1}(R,r,r')Z_{k_2}(R,r,r')
\eeq
Explicitly, in terms of pomeron Green functions and emission vertices,
one has
\[
I(k_1,k_2)=
\int\,d^{2}R \int d^{2}r \int d^{2}r'
\rho_p(r)\rho_q(r')
\int d^2R_1d^2R_2d^2z_1d^2z_2\]\beq
G_{s_1}(R+R_1,r|0,z_1)V_{k_1}(z_1)G_{s/s_1}(0,z_1|R_1,r')
G_{s_2}(R+R_2,r|0,z_2)V_{k_2}(z_2)G_{s/s_2}(0,z_2|R_2,r')
\eeq
By comparing this expression with the standard form of the double hard
scattering cross-section one may identify the double partonic density
of the projectile 
\beq
F_p(x_1,R_1,z_1|x_2,R_2,z_2)=g^4
\int d^{2}r \rho_p(r)G_{s_1}(R_1,r|0,z_1)G_{s_2}(R_2,r|0,z_2)
\eeq
where $\log s_{1(2)}=-\log x_{1(2)}$. A similar formula holds for the
target. The density (59) can be further integrated over $(1/2(R_1+R_2)$
to obtain the double parton distribution depending only on the distance
between the partons $\Gamma_p(x_1,x_2,R_1-R_2,z_1,z_2)$ defined by (21).

The expression (59) is a direct generalization of the single
parton density (43) and it can be obviously generalized to the multiple
partonic densities generated by the direct coupling of any number of pomerons
to the projectile (target)
\beq
F(x_1,R_1,z_1|...|x_n,R_n,z_n)=g^{2n}
\int d^{2}r \rho_p(r)
\prod_{i=1}^nG_{s_i}(R_i,r|0,z_i),\ \ \log x_i=-\log s_i
\eeq

In this simple picture the multiple partonic distributions are factorized
under the integral over $r$, which labels the different configurations of the
$q\bar{q}$ pair. Interestingly this is precisely the correlation in transverse
space which has been recently suggested\cite{ct} to describe the anomalously small value
of the effective cross section measured by the CDF experiment\cite{cdf}.

An estimate of the many-body parton distribution can be made, in this case, by
using the asymptotics (42)
and by substituting $r$ in all  slowly varying (logarithmic) factors
by its (small) average value $\langle r\rangle$. One obtains
\beq
F_p(x_1,R_1,z_1|...x_n,R_n,z_n)\simeq
\frac{\langle r^n\rangle}{\langle r\rangle ^n}
\prod_{i=1}^nF_p(x_i,R_i,z_i)
\eeq 
where $F_p(x,R,z)$ is the single parton distribution defined by (41).
Although the multi-parton distributions are just
a product of single ones, they contain the factor
$\langle r^n\rangle/\langle r\rangle ^n$ which is always greater than unity.
The consequence is a positive correlations between partons, which however
is independent of their position in transverse space. As an example
the factor 
$\langle r^2\rangle/\langle r\rangle ^2$ is equal to $4/\pi$ for the
Gaussian
distribution $\rho(r)$ and 3/2 for the exponential one.

\subsection{Triple pomeron: generalities}
\begin{center}
  \begin{figure}
    \includegraphics[width=0.6\textwidth,clip]{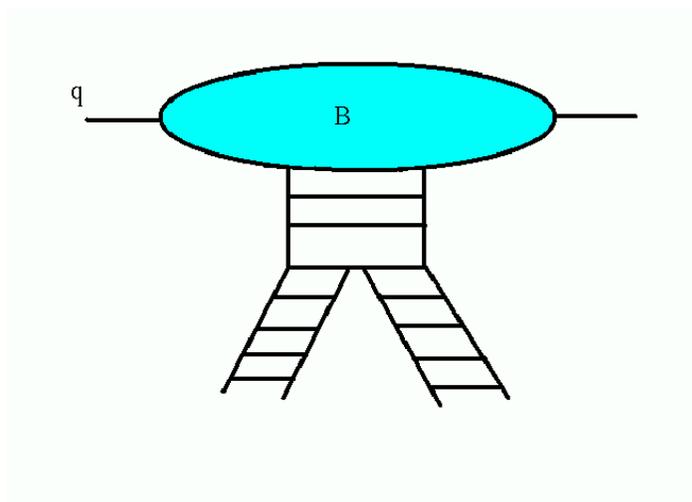}
    \caption{Double parton density in the triple pomeron case.}
    \label
{br5}
  \end{figure}
\end{center}

In Fig.\ref{br5} we show 
the amplitude in the case of triple
pomeron interaction which, in the large number of colors $N$
limit, has been discussed in ref.\cite{bv}. The resulting expression is
written as:
\[
{\cal A}_{TP}(s_1)=-is_1\frac{g^4N}{2\pi^3}\int_0^{y_1} dy
\int \frac{d^2r_1d^2r_2d^2r_3}{r_{13}^2r_{12}^2r_{23}^2}\]\beq
\Psi_1(r_1,r_2;y_1-y)\Psi_2(r_2,r_3;y_1-y)r_{13}^4\nabla_1^2\nabla_3^2
\Psi(r_1,r_3;y)
\eeq
Here and in the future we use the notations $r_{12}=r_1-r_2$,
$R_{12}=(r_1+r_2)/2$ etc. $y's$ are the corresponding rapidity
intervals.
The pomeronic amplitudes $\Psi$ are defined as follows
\beq
\Psi(r_1,r_3,y)=\frac{1}{2}g^2\int d^2R'_{13}d^2r'_{13}\rho(r'_{13})
G_{s_0}(r'_1,r'_3|r_1,r_3)
\eeq
with $s_0=e^y$ and similarly for the other two.
To obtain the corresponding double partonic density,
one has to substitute 1) the factor $-is_1$ by 2 and 2)  the two pomeron 
amplitudes $\Psi_1$ and $\Psi_2$
by the appropriate pomeronic Green functions with free legs describing 
the gluons taking  part
 in the hard interaction. The functions may
have different rapidity intervals,
corresponding to the different rapidities $y_1$ and $y_2$ of the interacting
gluons. The amplitude (62) is then converted into
\[
\frac{g^6N}{8\pi^3}\int_0^{y_{min}} 
dy\int d^2R'_{13}d^2r'_{13}\rho(r'_{13})\]\beq
\int \frac{d^2r_1d^2r_2d^2r_3}{r_{13}^2r_{12}^2r_{23}^2}
G_{s_1/s_0}(r_1,r_2|z_1,z_3)G_{s_2/s_0}(r_2,r_3|z_2,z_4)
r_{13}^4\nabla_1^2\nabla_3^2
G_{s_0}(r'_1r'_3|r_1,r_3)
\eeq
where $y=\log s_0$, $y_{min}={\rm min}\{y_1,y_2\}$.
 
Before moving further, one has to eliminate the differential operator
acting on the last Green function, which makes 
the incoming pomeron unsymmetric with respect to the two outgoing ones in the
triple vertex. To this purpose it is
sufficient to note that the differential operator is proportional to the Casimir operator of the
conformal group
\[
r^4_{13}\nabla_1^2\nabla_2^2=16C_2
\]
Acting on a function
\[ \left(\frac{r_{13}}{r_{10}r_{30}}\right)^{1+2i\nu}\]
which appears in $G(r'_1,r'_3|r_1,r_3)$ it gives factor
$16(\nu^2+1/4)^2$. The action of the operator on $G$ is therefore to
transform it into a Green function
which lacks the denominator in (39):
\beq
\tilde{G}_s(R,r,r')=\frac{1}{\pi^4}\int
d\nu\nu^{2}s^{-E(\nu)}
\int d^{2}r_{0}
\left(\frac{r}{r_{10}r_{20}}\right)^{1+2i\nu}
\left(\frac{r'}{r'_{10}r'_{20}}\right)^{1-2i\nu}
\eeq
Note that this function has the same asymptotics as $G$ at high $s$,
since it is determined by values of $\nu$ close to zero.

To obtain the double partonic density one has to finally take $Z_{13}=Z_{24}=0$,
and to fix the c.m. coordinates of the initial gluons in both pomeronic legs
$R_{12}\equiv R_{1}$ and $R_{23}\equiv R_2$.  Note that
the triple vertex coordinates $r_i$, $i=1,2,3$ can be expressed via the
c.m. coordinates of the initial gluons
\beq
r_1=R_{12}+R_{31}-R_{23},\ \  r_2=R_{12}+R_{23}-R_{13},
\ \ r_3=R_{23}+R_{13}-R_{12}
\eeq
so that
\beq
r_{12}=2(R_{13}-R_{23}),\ \  r_{23}=2(R_{12}-R_{13}),
\ \ r_{31}=2(R_{23}-R_{12})
\eeq
The Jacobian of the transformation from $r_1,r_2,r_3$ to 
$R_{12},R_{23},R_{31}$ is 4. To obtain the density one has to drop the
integrations over
$R_{12}$ and $R_{23}$. The double parton density
in the triple pomeron interaction case is therefore
\begin{eqnarray}
F(x_1,R_1,z_1|x_2,R_2,z_2)=&&
\frac{g^6N}{128\pi^3}
\frac{1}{(R_1-R_2)^2}\int_0^{y_{min}} dy
\int d^2r'
\rho(r')\nonumber\\
\int \frac{d^2R'd^2R}{(R-R_1)^2(R-R_2)^2}
&G&_{s_1/s_0}(R_1,2(R_2-R),z_1)\nonumber\\
\times&G&_{s_2/s_0}(R_2,2(R_1-R),z_2)
\tilde{G}_{s_0}(R'-R,r',2(R_{1}-R_{2}))
\end{eqnarray}
One can decouple the space
integration involving the
upper pomeron 
by shifting the integration variable $R'\rightarrow R'-R$. The expression is
then written as:
\begin{eqnarray}
F(&x&_1,R_1,z_1|x_2,R_2,z_2)=
\frac{g^6N}{128\pi^3}
\frac{1}{(R_1-R_2)^2}\nonumber\\
&\times&\int_0^{y_{min}}dy
\int d^2R'd^2r'
\rho(r')\tilde{G}_{s_0}(R',r',2(R_{1}-R_{2}))\nonumber\\
&\times&\int \frac{d^2R}{(R-R_1)^2(R-R_2)^2}
G_{s_1/s_0}(R_1,2(R_2-R),z_1)G_{s_2/s_0}(R_2,2(R_1-R),z_2)
\end{eqnarray}

\subsection{Triple pomeron: calculation}
To evaluate the expression (69) one needs to make some simplifications:
We use the asymptotic expressions for the two lower pomerons and we
keep explicitly into account that the two relative distances $z_{1,2}$ are
much smaller as compared to the other distanced in (69), as a consequence of
the large transverse momentum of the produced jets. A further simplification
is that the upper pomeron, although not
in its asymptotic regime, is taken with zero total gluon momenta,
where the Green function is substantially simpler.

We first discuss the integration on $R$
(third line in (69)):
\beq
I_1(y)=
\int \frac{d^2R}{(R-R_1)^2(R-R_2)^2}
G_{s_1/s_0}(R_1,2(R_2-R),z_1)G_{s_2/s_0}(R_2,2(R_1-R),z_2)
\eeq
The asymptotics of the two Green functions at large $s_{1,2}/s_0$ and
small $z_{1,2}$ is easily worked out (see Appendix). The resulting expression
is the same 
as in (42) with an additional factor 1/2 and a different definition of $\xi$:
\beq
G_{s_1/s_0}(R_1,2(R_2-R),z_1)_{z_1<<1}\simeq
 \frac{e^{\Delta\tilde{y}_1}}
{2\pi^{5/2}(a\tilde{y}_1)^{3/2}}\xi_1^{-1}
\ln \xi_1\exp \left(-\frac{\ln^{2} \xi_1}{a\tilde{y}_1}\right)
\eeq
where
\[ \xi_1=\frac{|R-R_1-R_2||R+R_1-R_2|}{2|R-R_2|z_1}, \ \ 
\tilde{y_1}=y_1-y\]
The same asymptotic expression holds for $G_{s_2/s_0}$, with 
$1\leftrightarrow 2$.

In the region $y_{1,2}>>y>>1$ one may neglect the dependence
on $y$ in all smoothly varying factors and one may substitute $\tilde{y}_{1,2}
\rightarrow y_{1,2}$ everywhere, with the exception of the exponent.
By shifting the integration variable $R\rightarrow R-R_1-R_2$ one finds
\[
I_1(y)=e^{-2\Delta y}e^{\Delta(y_1+y_2)}
\frac{z_1z_2}
{\pi^5a^3(y_1y_2)^{3/2}}\int \frac{d^2R}{R^2|R+R_1||R+R_2||R+2R_1||R+2R_2|}
\]\beq
\ln \xi_1\ln \xi_2\exp 
\left(-\frac{\ln^{2} \xi_1}{ay_1}-\frac{\ln^{2} \xi_2}{ay_2}\right)
\eeq
where
\beq
\xi_1=\frac{R|R+2R_1|}{2z_1|R+R_1|},\ \ \xi_2=\xi_1(1\rightarrow 2)
\eeq
The integral (72) is dominated by the small $R$ region, 
such that $\log(1/ R)\sim\sqrt{y_{1,2}}$. One then is allowed to put $R=0$ in all
factors which are finite in the $R=0$ limit and Eq.(72) is simplified to
\begin{eqnarray}
I_1(y)&=&e^{-2\Delta y}e^{\Delta(y_1+y_2)}
\frac{1}{4\pi^5a^3(y_1y_2)^{3/2}}
\frac{z_1z_2}{R_1^2R_2^2}\nonumber\\
&\times&
\int \frac{d^2R}{R^2}
\ln (R/z_1)\ln (R/z_2)\exp
 \left(-\frac{\ln^{2}(R/z_1)}{ay_1}-\frac{\ln^{2} (R/z_2)}{ay_2}\right)
\end{eqnarray}
The integration over $R$ now can be done explicitly (see Appendix) 
with the result
\[
I_1(y)=e^{-2\Delta y}e^{\Delta(y_1+y_2)}
\frac{\sqrt{\pi}}{4\pi^4}\left(\frac{1}{a (y_1+y_2)}\right)^{3/2}
\frac{z_1z_2}{R_1^2R_2^2}\]\beq
\left(1-\frac{2}{a(y_1+y_2)}\ln^2\frac{z_1}{z_2}\right)
\exp\left(-\frac{\ln^2(z_1/z_2)}{a(y_1+y_2)}\right)
\eeq
At large $y_1$ and $y_2$, with $y_1={\cal O}(y_2)$, the second term is sub-dominant
and it may be dropped. The expression is therefore reduced to
\beq
I_1(y)=e^{-2\Delta y}e^{\Delta(y_1+y_2)}
\frac{\sqrt{\pi}}{4\pi^4}\left(\frac{1}{a (y_1+y_2)}\right)^{3/2}
\frac{z_1z_2}{R_1^2R_2^2}
\exp\left(-\frac{\ln^2(z_1/z_2)}{a(y_1+y_2)}\right)
\eeq

The next step to evaluate (69) is to integrate the forward Green function, appearing in the
first line, over its rapidity variable with the factor
arising from $I_1$:
\beq
I_2=\int_0^{\infty}dye^{-2\Delta y}
\tilde{G}_{s_0,Q_{13}=0}(r',2(R_1-R_2))
\eeq
Here it has been explicitly indicated that the Green function has to be taken at
total momentum of the gluons equal to zero. One then obtains\cite{lipatov}
\beq
\tilde{G}_{s}(0,r',r)=\frac{rr'}{2\pi^2}\int_{-\infty}^{\infty}d\nu
s^{-E(\nu)}(r/r')^{2i\nu}
\eeq
where $E(\nu)$ is given by (40). 
By putting this expression into (77) and by integrating over $y$
one is left with an integral over $\nu$
\beq
I_2=\frac{rr'}{2\pi^2}\int d\nu (r/r')^{2i\nu}\frac{1}{2\Delta+E(\nu)}
\eeq
where $r=2(R_1-R_2)$. The distance $r'$ is of the oder of the projectile 
dimension and it is small, so that the ratio $r/r'$ for fixed $R_1-R_2$
is a large number. The integral (79) can be therefore evaluated by taking the residues 
of the integrand at the zeros of $2\Delta+E(\nu)$ in the upper half-plane.

The zeros are located at the points $\nu=ix_k$, $x_1<x_2,<...$. The first 
three points are (see\cite{b2}) 
\beq 
x_{1}=0.3169,\ \ x_{2}=1.3718,\ \ x_{3}=2.3867
\eeq
One obtains
\beq
I_2=\frac{1}{\alpha_s N}rr'\sum_kc_k(r/r')^{-2x_k}
\eeq
where
\beq
c_k^{-1}=\psi'(1/2-x_k)-\psi'(1/2+x_k)
\eeq
At large values of $r/r'$ the nearest pole contributes, so that one finds
\beq
I_2\simeq\frac{1}{\alpha_sN}c_1rr'(r/r')^{-2x_1}
\eeq

The final average, namely the integration on $r'$ with the color density of
the projectile, gives
therefore 
$
\langle r^{1.64}\rangle
$, to be compared with $
\langle r^2\rangle
$ which is the result obtained for the 
double parton density when coupling the two pomerons directly to the
$q{\bar q}$ pair.

Collecting all the factors the double parton density
corresponding to the triple pomeron picture is finally written as
\[
F(x_1,R_1,z_1|x_2,R_2,z_2)\]\beq=
\frac{c_1}{4^{1+x_1}\pi^{7/2}}\alpha_s^2
\langle r^{1+2x_1}\rangle
e^{\Delta(y_1+y_2)}
\left(\frac{1}{a(y_1+y_2)}\right)^{3/2}
\frac{z_1z_2}{R_1^2R_2^2|R_1-R_2|^{1+2x_1}}
\exp\left(-\frac{\ln^2(z_1/z_2)}{a(y_1+y_2)}\right)
\eeq 
Analogously to the case
of direct coupling of the pomerons to the $q{\bar q}$ pair,
the double parton density contains the factor $1/R_1^2R_2^2$. However in the
triple pomeron case the factor $1/|R_1-R_2|^{1+2x_1}$ induces 
an additional strong positive correlation in transverse space between the two
partons. 

\section{Conclusions}

In the present paper we have evaluated the double parton distributions, in the
case of interactions between very virtual $q{\bar q}$ pairs in the
BFKL regime. After factorizing the
hard interactions, the double parton
distributions depend on the structure of the blob attached to the
projectile (or to the target). Two limiting possibilities have therefore been
considered: $a$) the 
whole interaction is represented by the exchange of two BFKL pomerons attached
directly to the $q{\bar q}$ pair. $b$) The process is described by triple
pomeron interaction, where the blob is itself represented by a BFKL
pomeron. The two-body correlation in the double distribution are different in
the two cases. 
In the first case the only correlation in the double parton distribution is
induced by the configuration taken by the 
$q{\bar q}$ pair in transverse space, namely at a given $q{\bar q}$
configuration the resulting many-body
parton distribution is just a Poissonian. The second case is, on the contrary,
characterized by a strong 
correlation in transverse space, as the common source of the two partons is a BFKL ladder rather than the
$q{\bar q}$ pair. The two possibilities are
linked, through the AGK rules, to the diffractive events and to the
events with rapidity gaps. The relative importance of the two 
contributions can
be therefore inferred by comparing the rates of the two processes.

\vskip.2in

{\bf APPENDIX}
\vskip.2in
In this appendix we shall give some technical details concerning the 
asymptotics (79) and calculation of the integral in (72).

First the asymptotics of $G_s(R,r,r')$ at small $r'$. From (37) one
concludes that at $s\rightarrow\infty$ small values of $\nu$ contribute.
So one is lead to study the leading behavior of the integral over $r_0$
at small $\nu$ when $r'\rightarrow 0$. In this limit the second factor in
(37) is singular at $r_0=0$. So small values of $r_0$ give the dominant
contribution and one can take the first factor out of the integral over
$r_0$ at $r_0=0$. The integral over $r_0$ then simplifies to
\beq
I=\int\frac{d^2r_0}{(r_0|r'-r|)^{1-2i\nu}}
\eeq
(To keep the integral convergent one should take ${\rm Im}\,\nu>0$).
This integral is trivially calculated by going over to the momentum
space:
if
\beq r^{-1+2i\nu}=\int\frac{d^2q}{(2\pi)^2}e^{iqr}f(q) \eeq
then
\beq
I=\int\frac{d^2q}{(2\pi)^2}e^{iqr'}f^2(q)
\eeq
Function $f(q)$ is trivially found by the inverse Fourier transformation:
\beq
f(q)=\pi 2^{1+2i\nu}q^{-1-2i\nu}\frac{\Gamma(1/2+i\nu)}{\Gamma(1/2-i\nu)}
\eeq
Putting this into (83) and doing the integration we find
\beq
I=(i\pi/2){r'}^{4i\nu}\frac{1}{\nu}
\eeq
Inserting this expression into the integral over $\nu$ together with
all accompanying factors and taking its asymptotics at $s\rightarrow\infty$
one obtains the desired asymptotics (79).

As to the integral which appears in (72), in the variable $\beta=\ln R$
it reduces to an integral of the Gaussian type
\beq
\int_{-\infty}^{+\infty}d\beta (\beta-\beta_1)(\beta-\beta_2)
\exp\left(-\frac{(\beta-\beta_1)^2}{ay_1}-\frac{(\beta-\beta_2)^2}{ay_2}
\right)
\eeq
which can be done in a straightforward manner, with a result (73).
\vskip.2in
{\bf Acknowledgments.}
\vskip.2in
M.A. Braun thanks the INFN and the University of Trieste
for their financial help during his stay
at Trieste.
This work was partially supported by the Ministero dell'Universit\`a e della
Ricerca Scientifica (MURST) through the funds COFIN99.



\end{document}